\documentclass[aps,prc,superscriptaddress,twoside,twocolumn,nofootinbib,%
showpacs,floatfix,dvips]{revtex4}

\usepackage{amsmath,amssymb,graphicx,bm}

\usepackage[normalem]{ulem} % \sout{old text} for strikeout
\usepackage[dvips]{color} % For blue in-text comments and additions
\renewcommand\sout{\bgroup \color{red} \ULdepth=-.5ex \ULset}

\begin{document}

%%%%%%%%%%%%%%%%%%%%% Title %%%%%%%%%%%%%%%%%%%%%%

\title{\boldmath
Deuteron production and elliptic flow in relativistic heavy ion collisions}

%%%%%%%%%%%%%%%%%%%% Authors %%%%%%%%%%%%%%%%%%%%%

\author{Yongseok Oh}%
\email{yoh@kisti.re.kr}

\affiliation{Cyclotron Institute and Physics Department, Texas A\&M
University, College Station, Texas 77843, USA}

\affiliation{Korea Institute of Science and Technology Information, Daejeon 305-806, Korea}

\author{Zi-Wei Lin}%
\email{linz@ecu.edu}

\affiliation{Department of Physics, East Carolina University,
Greenville, North Carolina 27858, USA}

\author{Che Ming Ko}%
\email{ko@comp.tamu.edu}

\affiliation{Cyclotron Institute and Physics Department,
Texas A\&M University, College Station, Texas 77843, USA}

\date{\today}

%%%%%%%%%%%%%%%%%%%% Abstract %%%%%%%%%%%%%%%%%%%%%

\begin{abstract}

The hadronic transport model \textsc{art} is extended to include the production and annihilation of deuterons via the reactions $BB \leftrightarrow dM$, where $B$ and $M$ stand for baryons and mesons, respectively, as well as their elastic scattering with mesons and baryons in the hadronic matter. This new hadronic transport model is then used to study the transverse momentum spectrum and elliptic flow of deuterons in relativistic heavy ion collisions, with the initial hadron distributions after hadronization of produced quark-gluon plasma taken from a blast wave model. The results are compared with those measured by the PHENIX and STAR Collaborations for Au+Au collisions at $\sqrt{s_{NN}^{}} = 200$~GeV, and also with those obtained from the coalescence model based on freeze-out nucleons in the transport model.

\end{abstract}


\maketitle

\section{Introduction}

An important observable in heavy ion collisions at the Relativistic Heavy Ion Collider (RHIC) is the azimuthal anisotropy of the momentum distributions of produced particles in the plane perpendicular to the beam direction, particularly the so-called elliptic flow ($v_2^{}$) that corresponds to the second Fourier coefficient in their azimuthal angle distribution~\cite{Olli92,VZ94}. The measured elliptic flow is not only large but also shows a constituent quark number scaling, especially at intermediate transverse momenta, i.e., the dependence of the elliptic flows of identified hadrons on their transverse momenta becomes similar if both are divided by the number of constituent quarks in a hadron. This scaling behavior of hadron elliptic flows is well described by the quark coalescence model for hadron production from the quark-gluon plasma (QGP) formed in relativistic heavy-ion collisions~\cite{GKL03,GKL03b,HY03,FMNB03,MV03,KCGK04}. Also, the measured elliptic flows of identified hadrons follow a mass ordering at low transverse momenta, namely, the strength of the elliptic flow becomes smaller as the hadron mass increases. This has
also been well described by the transport model~\cite{LK02} as well as by the ideal hydrodynamics~\cite{KSH00,HKHRV01}.

Recently, the elliptic flow of deuterons has been measured in Au+Au collisions at energy $\sqrt{s_{NN}^{}} = 200$~GeV~\cite{PHENIX07,STAR07a}. The data from the PHENIX Collaboration covers the intermediate transverse momentum ($p_T^{}$) region~\cite{PHENIX07}, while the STAR Collaboration has made measurements in a wider range of $p_T^{}$ including the low $p_T^{}$ region~\cite{STAR07a}. The two measurements agree well at intermediate $p_T^{}$ region ($p_T^{} > 1.5$~GeV$/c$) except that the STAR data show a negative elliptic flow at low $p_T^{}$ ($p_T^{} < 1$~GeV$/c$). A negative elliptic flow has also been seen in the preliminary data from the PHENIX Collaboration for $J/\psi$'s at $p_T^{} \sim 1.5$~GeV$/c$~\cite{PHENIX08,PHENIX08a}. Negative values of deuteron elliptic flow cannot be explained by the coalescence model unless nucleons have negative elliptic flows~\cite{OK07}, while negative $J/\psi$ flows may~\cite{KB08} or may not~\cite{LIN09} require negative charm quark elliptic flows. Since measured nucleon elliptic flow does not show any negative value in small transverse momentum region~\cite{PHENIX07,STAR08}, observed negative values of deuteron elliptic flow raise an interesting question on the mechanism for their production and interactions in relativistic heavy ion collisions.

In Ref.~\cite{OK07}, two of us have studied deuteron production at RHIC in a dynamical model which is based on the time-dependent perturbation theory. Using the elementary reactions of $NN \to d\pi$, $NNN \to dN$, and $NN\pi \to d\pi$, we have computed the production rate of deuterons using the measured nucleon transverse momentum distribution that is parameterized by an effective temperature and a momentum-dependent elliptic flow. Although the energy is conserved in this approach, in contrast with the coalescence model, it needs the introduction of the reaction time and volume as parameters to fix the multiplicity of deuterons. The resulting deuteron $p_T^{}$ spectrum and elliptic flow are found to be similar to those of the coalescence model based on the same nucleon momentum distribution. Compared to the experimental data~\cite{PHENIX07,STAR07a}, the calculated deuteron $p_T^{}$ spectrum is, however, too soft. As discussed in Ref.~\cite{OK07}, this is due to the use of an effective temperature to model the effect of radial flow, resulting in the absence of correlations between nucleon positions and momenta. For the elliptic flow of deuterons, this approach describes reasonably the experimental data except at small $p_T^{}$, where it gives positive values while the data show negative values~\cite{STAR07a}.

To overcome the shortcomings of the model in Ref.~\cite{OK07}, we use in the present work a transport model to study deuteron production and elliptic flow in relativistic heavy ion collisions. We assume that the particles produced from hadronization of the created quark-gluon plasma are in thermal and chemical equilibrium and undergo collective motions, similar to those described by the blast wave model~\cite{SSH93}, and they then interact with each other via hadronic rescattering. For the latter, we use a relativistic transport model (\textsc{art})~\cite{LK95,LK96,LSZK01} code embedded in a multiphase transport model (\textsc{ampt})~\cite{LKLZP04}. We also carry out a coalescence model calculation using the freeze-out nucleons to form deuterons. This approach is different from the coalescence model used in Ref.~\cite{OK07} as it includes the position and momentum correlations of final nucleons through their collective motions introduced in  the initial blast wave model and further generated by final hadronic rescattering.

This paper is organized as follows. In the next section, we describe the blast wave model used for generating the hadrons produced at hadronization that are taken as the initial conditions in the transport model for describing hadronic rescattering. Section~\ref{sec:hadrons} gives the results for the transverse momentum distributions and elliptic flows of pions and protons after hadronic scattering. The results for the deuteron transverse momentum distribution and elliptic flow are presented and discussed in Sec.~\ref{sec:deuteron1}.  These results are further compared in Section~\ref{sec:coal} with those from the dynamical coalescence model. We also compare in Section~\ref{sec:emission} the deuteron emission time distributions from these two models and discuss in Section~\ref{sec:scaling} the nucleon number of scaling of deuteron elliptic flow.
Section~\ref{sec:summary} contains a summary and discussions. The elementary production and annihilation processes for deuterons that are included in the transport model are described in the Appendix.

\section{Initial distributions of hadrons} \label{sec:initial}

In relativistic heavy ion collisions, where a quark-gluon plasma is produced, hadrons are initially formed during hadronization of the QGP. To describe the distribution of these hadrons, we use the blast-wave model which assumes that they are in thermal and chemical equilibrium and undergo a collective expansion. As in the \textsc{art} model, we include in the present work mesons such as $\pi$, $\rho$, $\omega$, $\eta$, $K$, $K^*$, and $\phi$ and baryons such as $N$, $\Delta$, $\Lambda$, and $\Sigma$ as well as their anti-particles. We also include both deuterons and anti-deuterons in the initial distributions. The Lorentz-invariant thermal distribution $f(x,p)$ of these particles at temperature $T_C^{}$ is then given by
\begin{equation}
f(x,p) \propto \exp\{-p^\mu u_\mu/T_C^{}\},
\end{equation}
where the four-momentum $p^\mu$ is
\begin{equation}
p^\mu = (p^0,\bm{p}) = (m_T^{}\cosh y, p_T^{} \cos\phi_p,
p_T^{}\sin\phi_p, m_T^{} \sinh y) \label{eq:momentum}
\end{equation}
and the flow four-velocity $u_\mu(x)$ is
\begin{eqnarray}
u^\mu &=& \cosh\rho ( \cosh\eta, \tanh\rho \cos\phi, \tanh\rho
\sin\phi, \sinh\eta) \nonumber \\
&=& \gamma_T^{} (\cosh\eta, \bm{\beta}, \sinh\eta).
\end{eqnarray}
In the above, $y$ is the energy-momentum rapidity, $m_T^{}$ ($= \sqrt{m^2 + p_T^2}$) is the transverse mass with $m$ being the mass of the considered particle, $\gamma_T^{} = 1/\sqrt{1-\beta^2}$, and $\eta$ ($\rho$) is the longitudinal (transverse) flow rapidity. We have followed the usual convention to define the $z$-axis along the beam direction and the $x$-axis in the reaction plane of the collision. The angles $\phi_p$ and $\phi$ in the above are the azimuthal angles of the momentum and position vectors of a particle with respect to the $x$-axis. For the longitudinal flow, it is assumed to be boost invariant so that the longitudinal flow rapidity is equal to the energy-momentum rapidity, i.e., $\eta=y$. We further assume that the distribution in the energy-momentum rapidity is uniform in the midrapidity.

Because of non-vanishing elliptic flow $v_2^{}$ in non-central heavy ion collisions, which is defined as
\begin{equation} v_2^{} =
\left\langle \frac{p_x^2 - p_y^2}{p_x^2 + p_y^2} \right\rangle
\end{equation}
with $p_x$ and $p_y$ being, respectively, the projections of the particle transverse momentum along the $x$ and $y$ axes in the transverse plane, the transverse flow velocity is anisotropic with respect to the azimuthal angle $\varphi$. We thus parameterize the transverse flow velocity as
\begin{equation}
\bm{\beta} = \beta(r) \left[ 1 + \varepsilon(p_T^{}) \cos(2\varphi)
\right] \hat{\bm{n}},
\label{eq:beta}
\end{equation}
where $\hat{\bm{n}}$ is the unit vector in the direction of their transverse flow velocity $\bm{\beta}$, which is taken to be normal to the surface of the hadronic system to be defined below. We have also introduced a $p_T^{}$-dependent coefficient $\varepsilon$ to model the saturation of the resulting $v_2^{}$ at large $p_T^{}$ as observed in experiments. Specifically, we parameterize $\varepsilon$ as
\begin{equation}
\varepsilon(p_T^{}) = c_1^{} \exp(-p_T^{}/c_2^{}).
\end{equation}
We further parameterize the radial flow velocity as
\begin{equation}
\beta(r) = \beta_0 \left( \frac{r}{R_0} \right),
\end{equation}
where $R_0$ is a parameter related to the transverse size of the initial hadron distribution in space and $r$ is the distance of the particle from the origin of the coordinate system in the transverse plane.

For the spatial distribution of the particles produced at hadronization, it is assumed to be inside a cylinder with its axis along the longitudinal direction and having an elliptic shape in the transverse plane, as only minimum biased collisions are considered in the present study. These particles are further assumed to be uniformly distributed in the transverse plane. This means that the initial hadrons are uniformly distributed in the spatial region of
\begin{equation}
\left( \frac{x}{A} \right)^2 + \left( \frac{y}{B} \right)^2 \le 1.
\label{eq:ellipse1}
\end{equation}
In terms of the spatial elliptic anisotropy~\cite{LK02}
\begin{equation}
s_2^{} = \left\langle\frac{x^2 - y^2}{x^2 + y^2}\right\rangle,
\end{equation}
the spatial region as given by Eq.~(\ref{eq:ellipse1}) can be rewritten as
\begin{equation}
r \le R_0 \left[ 1 + s_2^{} \cos(2\varphi) \right],
\end{equation}
leading to
\begin{equation}
A = R_0 (1 + s_2^{}), \qquad B = R_0 (1 - s_2^{}).
\end{equation}
For the coordinate system defined above, the initial shape of the hadronic system in non-central collisions then has $s_2^{} \le 0$.

We further introduce a formation time $\tau_0^{}$ for hadrons produced from hadronization as in Ref.~\cite{LKLZP04}. Because of boost invariance, the time and the position in $z$-direction of an initial hadron are given by $t = \tau_0^{} \cosh y$ and $z = \tau_0^{} \sinh y$, where $y$ stands for rapidity.

Since we do not consider the dynamics of partons in QGP before hadronization, the number of hadrons in the initial state cannot be determined a priori. We thus treat the initial number of positively charged pions $N_{\pi^+}$ as a free parameter so that its number after hadronic rescattering is fitted to the measured pion multiplicity. The number of other hadrons after hadronization is then determined from the assumption that they are in thermal and relative chemical equilibrium with respect to pions.

For Au$+$Au collisions at center of mass energy $\sqrt{s_{NN}^{}} = 200$~GeV, the parameters for the initial conditions are taken as follows. For the spatial distribution of initial hadrons, we use the parameters
\begin{eqnarray}
R_0 = 5.0 \mbox{ fm}, \quad s_2^{} = -0.05, \quad \tau_0^{} =
2.5~\mbox{fm/c},
\end{eqnarray}
similar to the quark distributions obtained in the \textsc{ampt} model at hadronization. For the transverse momentum distribution, we use the critical temperature $T_C^{} = 175$~MeV and
\begin{eqnarray}
\beta_0 = 0.55, \quad c_1^{} = 0.43, \quad c_2^{} = 0.85 \mbox{ GeV}/c
\label{eq:parameters}
\end{eqnarray}
for the flow velocity parameterized in Eq.~(\ref{eq:beta}). The number of initial positively charged pions is taken to be $N_{\pi^+} = 100$. This then leads to a proton number of $N_p = 2.5$ in the thermal model with zero charge and baryon chemical potentials. These pions and protons as well as other hadrons are then uniformly distributed in the rapidity region $|y| < 2.0$. The resulting initial transverse momentum spectra and elliptic flows of midrapidity ($|y| \le 0.5$) $\pi^+$'s and protons, whose numbers are 71 and 6.4, respectively after including those from decays of meson and baryon resonances, are shown in Figs.~\ref{fig:pt-all} and \ref{fig:v2-all}. These results will be compared in the next section with those of the freeze-out pions and protons after hadronic scattering.

\section{Pion and nucleon transverse momentum spectra and elliptic flows}\label{sec:hadrons}

%%% FIG 1

\begin{figure}[t]
\centering
\includegraphics[width=0.85\columnwidth,angle=0,clip]{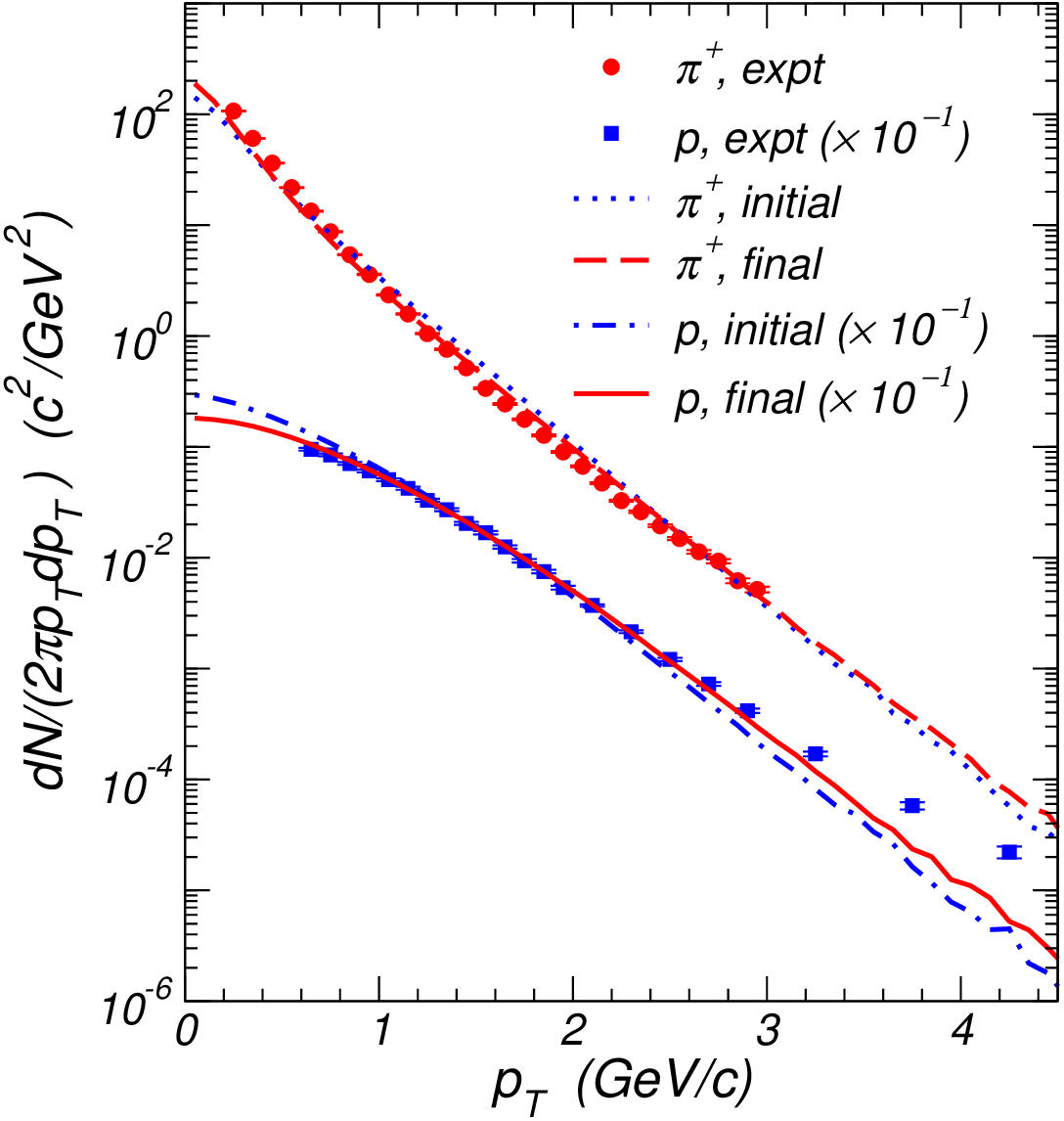}
\caption{(Color online) Transverse momentum spectra of pions and protons in midrapidity $|y| < 0.5$. Data are from Ref.~\cite{PHENIX03b}.}
\label{fig:pt-all}
\end{figure}

We first discuss the transverse momentum spectra of pions and protons after hadronic evolution. The results in the midrapidity region $|y| < 0.5$ are shown in Fig.~\ref{fig:pt-all} by the dashed and solid lines for pions and protons, respectively, and they are comparable to the experimental data measured by the PHENIX Collaboration~\cite{PHENIX03b}. It is seen that the final pion and proton transverse momentum spectra are not very different from the initial ones shown by the dotted and dash-dotted lines, respectively. This can be understood from the fact that the effect of decreasing temperature of the hadronic matter due to expansion is compensated by an increase in the collective flow velocity. A similar conclusion was drawn in Ref.~\cite{HHKLN07b}, where the effect of hadronic rescattering was studied through the hadronic transport model \textsc{jam}~\cite{NOONC99}. We note that the number of pions and protons in midrapidity after hadronic rescattering are $N(\pi^+) \approx 74$ and $N(p)\approx 5.4$, respectively, including those from decays of baryon and meson resonances, and are only slightly different from their initial values.

%%% FIG 2

\begin{figure}[t]
\centering
\includegraphics[width=0.95\columnwidth,angle=0,clip]{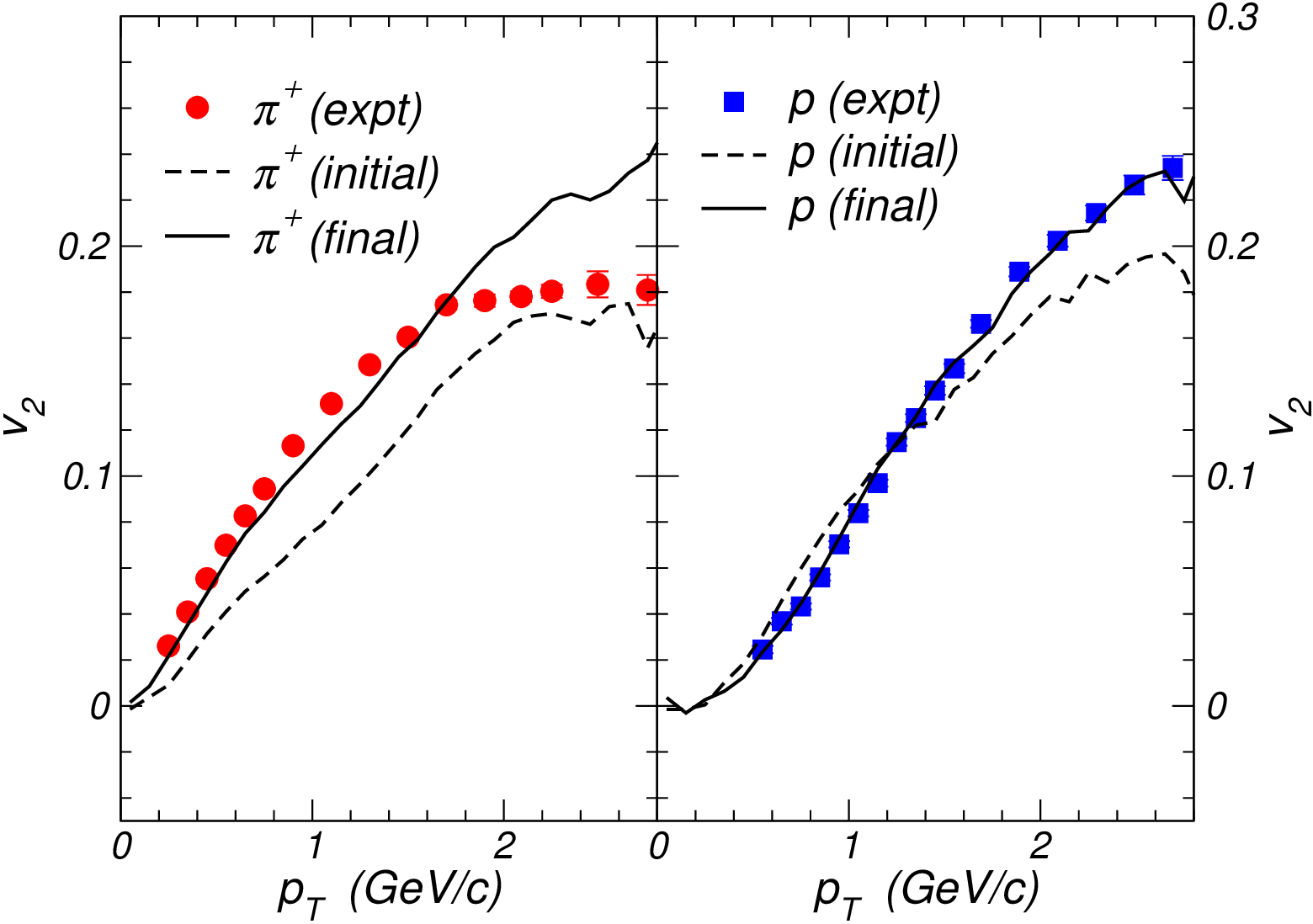}
\caption{(Color online) Elliptic flow $v_2^{}$ of pions and protons for rapidities $|y| < 0.5$. Data are from Ref.~\cite{PHENIX07}.}
\label{fig:v2-all}
\end{figure}

In Fig.~\ref{fig:v2-all}, the elliptic flow $v_2^{}$ of pions and protons are shown by solid lines as functions of transverse momentum. It is seen that our model describes reasonably the experimental data from the PHENIX Collaboration except for $p_T^{} > 2$~GeV$/c$, where it overestimates the measured  pion elliptic flow.  We have found that it is very difficult to find the parameters used for the elliptic flow in Eq.~(\ref{eq:beta}) that would describe both the pion and proton $v_2^{}$ in all $p_T^{}$ region. This may not be surprising as the pion and proton elliptic flow at high $p_T^{}$ follow the quark number scaling and are thus unlikely to be described by the blast wave model. To overcome this may require the use of the blast wave model for the quark distributions and the quark coalescence model to generate the initial meson and baryon distributions. In contrast to the transverse momentum spectra, the final pion and proton elliptic flows are different from the initial ones shown by dashed lines. One sees that the pion elliptic flow is increased by scattering. For protons, their elliptic flow after scattering also increases at large $p_T^{}$ but becomes smaller at low $p_T^{}$, reflecting an enhanced mass-ordering effect as a result of increasing radial flow.

\section{Deuteron transverse momentum spectrum and elliptic flow}
\label{sec:deuteron1}

\begin{figure}[t]
\centering
\includegraphics[width=0.85\columnwidth,angle=0,clip]{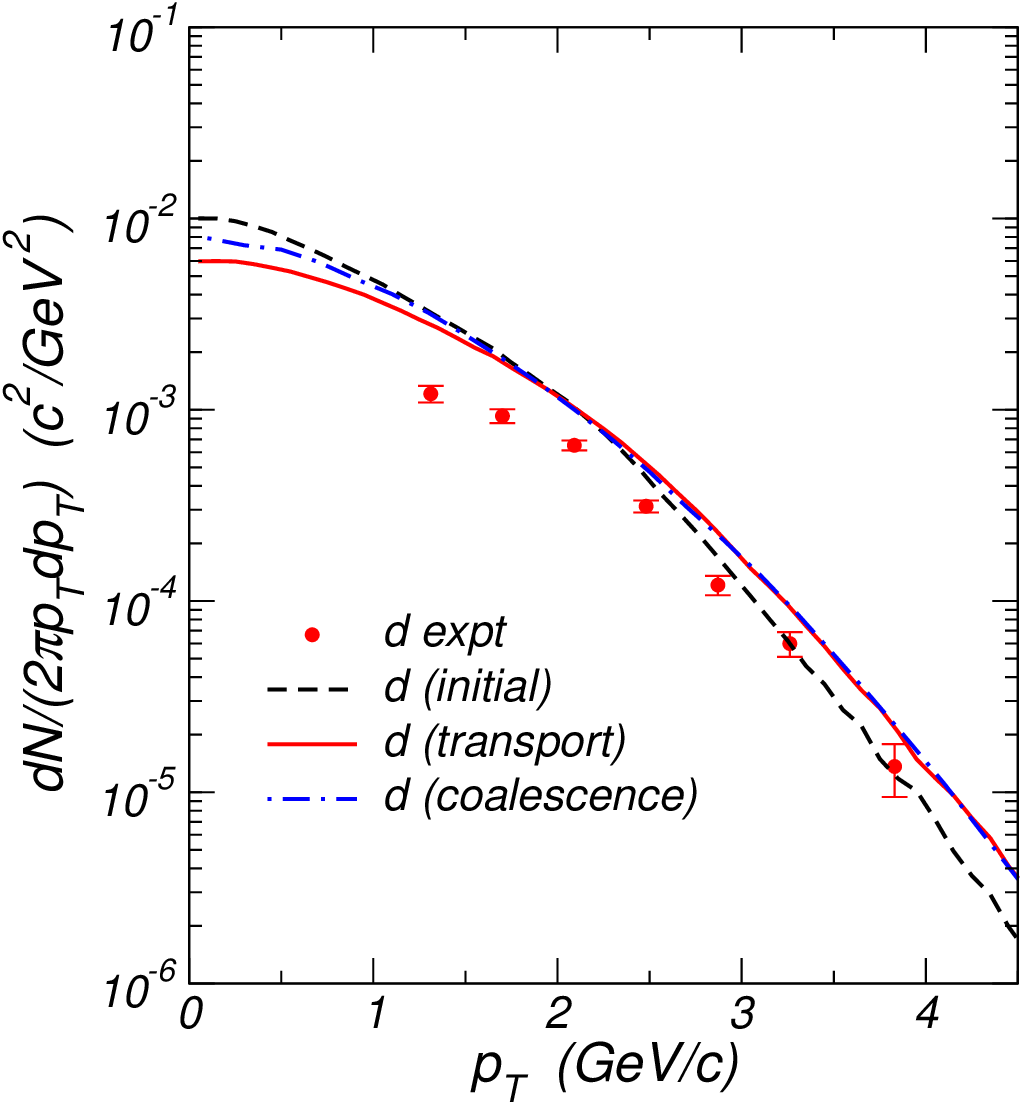}
\caption{(Color online) Transverse momentum spectrum of deuterons in midrapidity $|y| < 0.5$. Dashed and solid lines are for initial and final deuterons, while the dash-dotted line is from the coalescence model. Data are from Ref.~\cite{PHENIX05}.}
\label{fig:deut-pt}
\end{figure}

In this section, we study the $p_T^{}$ spectrum of deuterons and their elliptic flow. As for other hadrons, we have assumed that they are initially produced at hadronization and then undergo hadronic scattering through which they can be annihilated and also reproduced. To include these effects, we have modified the \textsc{art} code embedded in the \textsc{ampt} model as described in detail in the Appendix. The resulting deuteron $p_T^{}$ spectrum is given in Fig.~\ref{fig:deut-pt} by the solid line and is seen in relatively good agreement with the PHENIX data~\cite{PHENIX05}. The  calculated deuteron number of about $0.046$ in the midrapidity is, however, somewhat larger than the measured one. Compared with the deuteron initial $p_T^{}$ spectrum shown by the dashed line, the final one has a larger inverse slope reflecting the further development of the radial flow of deuterons by hadronic rescattering.

%%% FIG 3

\begin{figure}[t]
\centering
\includegraphics[width=0.85\columnwidth,angle=0,clip]{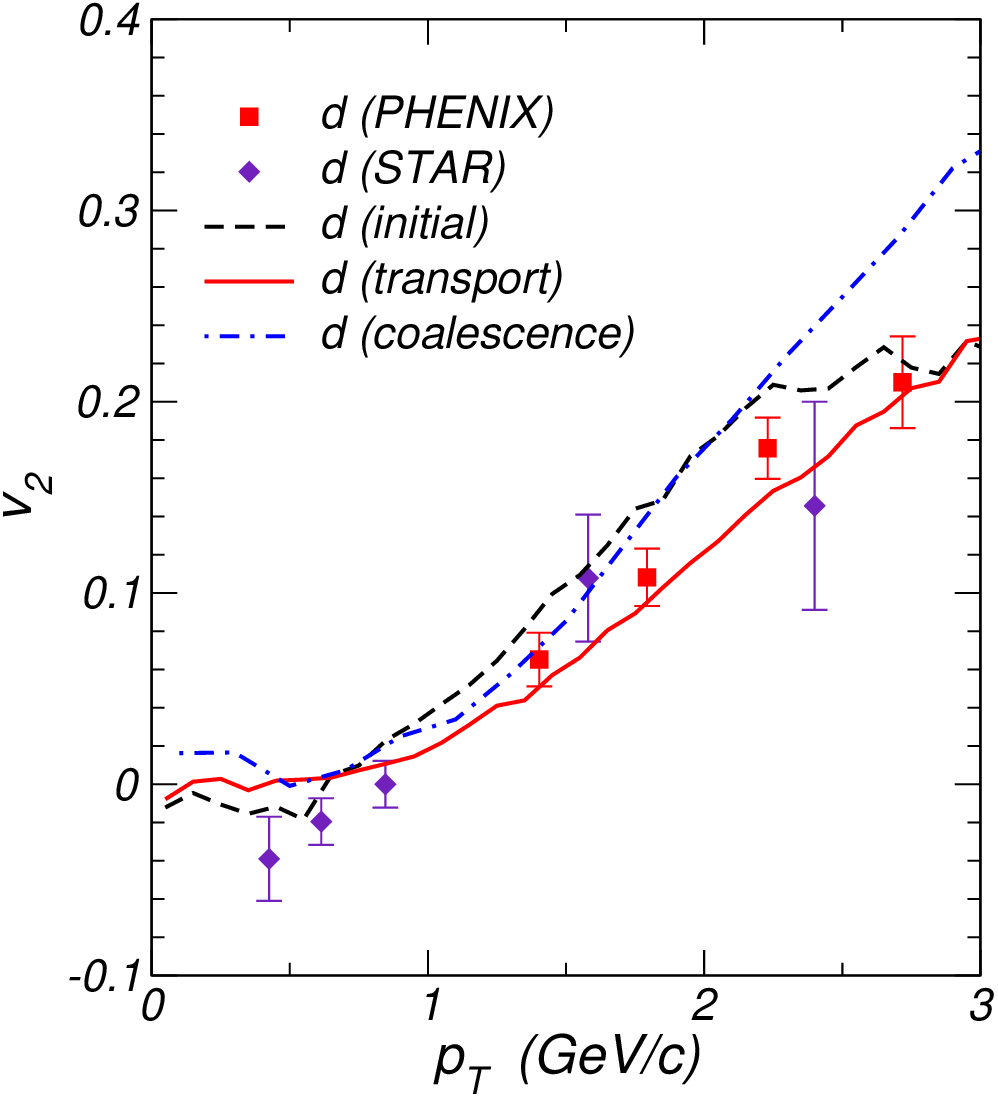}
\caption{(Color online) Same as Fig.~\ref{fig:deut-pt} for the deuteron elliptic flow. Data are from Refs.~\cite{PHENIX07,STAR07a}.} \label{fig:deut-v2}
\end{figure}

Shown in Fig.~\ref{fig:deut-v2} by the solid line is the calculated final deuteron elliptic flow, which is seen to describe reasonably the experimental data. The negative values of deuteron elliptic flow at low $p_T^{}$ ($< 1$~GeV$/c$) reported by the STAR Collaboration~\cite{STAR07a} are, however, not reproduced by the present model. We first note that the deuteron elliptic flow from the initial blast wave model shown by the dashed line does have small negative values at very low $p_T^{}$. This is due to following combined effects. Because of the larger flow velocity in the reaction plane than out of the reaction plane, deuterons of a given transverse momentum $p_T^{}$ in the fire-cylinder frame have a smaller transverse momentum in the local frame if they move in the reaction plane than if they move out of the reaction plane. Since the number of deuterons with transverse momentum $p^\prime_T$ in the local frame is proportional to $p^\prime_T \exp(-m^\prime_T/T)$, it is an increasing or decreasing function of $p^\prime_T$ for $p^\prime_T \lesssim \sqrt{m_d T}$ or $p^\prime_T \gtrsim \sqrt{m_d T}$.  While the resulting deuteron $v_2^{}$ is positive in the latter case as there are more deuterons of transverse momentum $p_T^{}$ moving in the reaction plane than moving out of the reaction plane, it is negative in the former case. This effect is, however, reduced during subsequent hadronic scattering as a result of increasing radial flow velocity and additional production of deuterons. Indeed, the final deuteron $v_2^{}$ at low $p_T^{}$ is  close to zero.  Similarly, the deuteron elliptic flow at larger $p_T^{}$ is reduced after hadronic evolution as shown in Fig.~\ref{fig:deut-v2}.

%%% FIG 4

\begin{figure}[t]
\centering
\includegraphics[width=0.75\columnwidth,angle=0,clip]{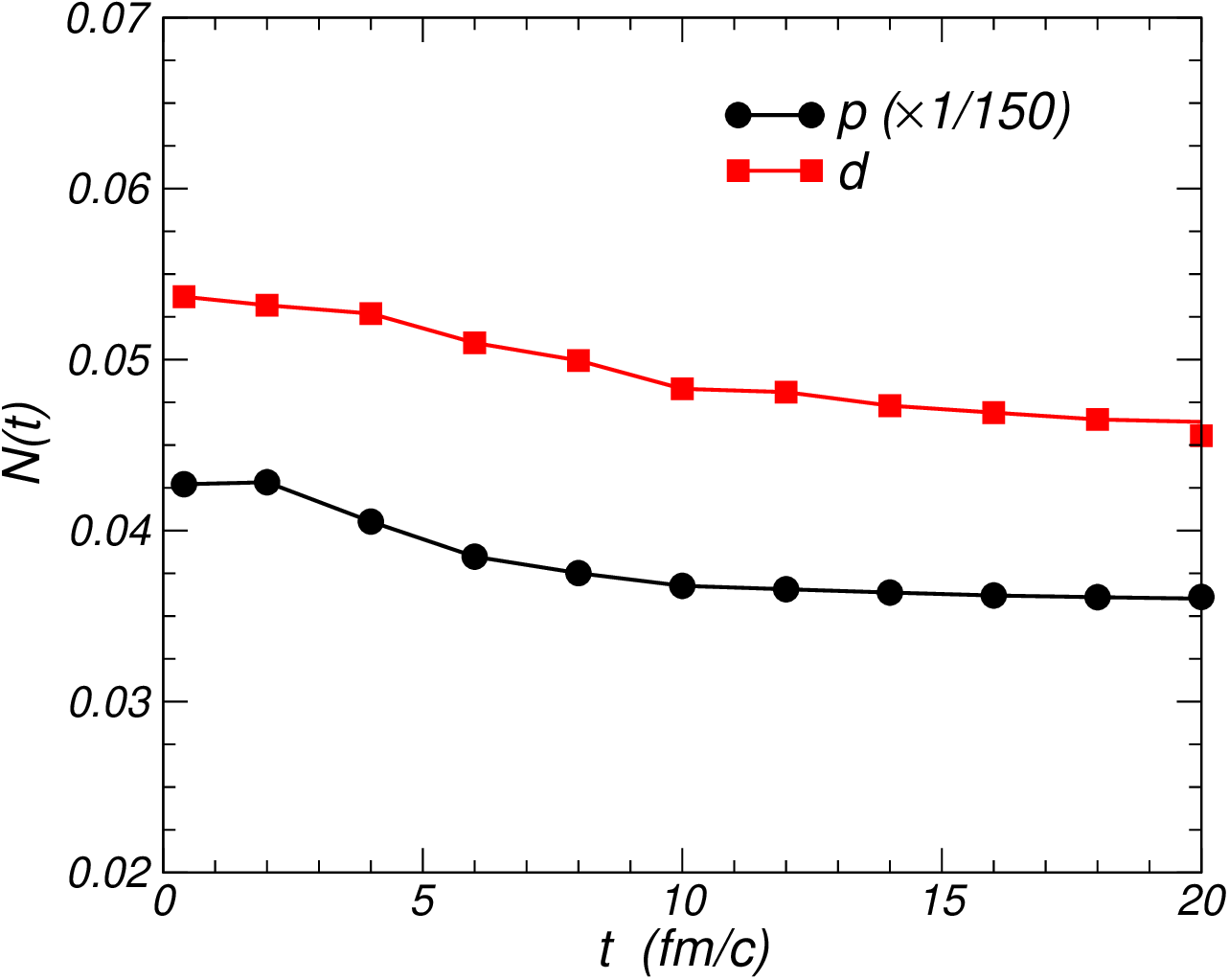}
\caption{(Color online) Proton (scaled by a factor of $1/150$) and deuteron numbers versus time in midrapidity $|y|< 0.5$.} \label{fig:number-time}
\end{figure}

In Fig.~\ref{fig:number-time}, we show the numbers of protons and deuterons in the midrapidity $|y|\le 0.5$ as functions of time. Both are seen to decrease gradually with time and this is partially due to scattering into larger rapidity regions.

%%% FIG 5

\section{The dynamical coalescence model}
\label{sec:coal}

It is of interest to compare the results from the transport model with those from the coalescence model. In the coalescence model, deuterons are produced by recombination of nucleons at freeze-out using the sudden approximation. The momentum distribution of produced deuterons in this model is given by~\cite{GKL03b,CKL03a}
\begin{eqnarray}
\frac{d^3N_d}{dp_d^3} &=& g \int d^3 x_1^{} d^3 x_2^{} d^3 p_1^{} d^3
p_2^{} \frac{d^6 N_p}{dx_1^3 dp_1^3} \frac{d^6 N_n}{dx_2^3 dp_2^3}
\nonumber \\
&& \mbox{} \times f_W(\bm{x}_1',\bm{x}_2';\bm{p}_1',\bm{p}_2')
\delta^{(3)}(\bm{p}_d^{} - \bm{p}_1^{} - \bm{p}_2^{})
\end{eqnarray}
with $g=3/4$ from the consideration of the spins of nucleons and deuterons. In the above, ${d^6 N_p}/{dx_1^3 dp_1^3}$ and ${d^6 N_n}/{dx_2^3 dp_2^3}$ are, respectively, the spatial and momentum distributions of protons and neutrons in the fire-cylinder frame and $f_W(\bm{x}_1',\bm{x}_2';\bm{p}_1',\bm{p}_2')$ is the Wigner function of the deuteron in its rest frame. Assuming that the wave function of the deuteron is given by that of a harmonic oscillator, its Wigner function is then
\begin{equation}
f_W(\bm{x}_1',\bm{x}_2';\bm{p}_1',\bm{p}_2') = 8 \exp\left( -
\bm{p}^2 \sigma^2 - \bm{x}^2/\sigma^2 \right),
\end{equation}
where $\sigma = 1/\sqrt{\mu \omega}$ with the reduced mass
$\mu = m_N^{}/2$ and
\begin{equation}
\bm{x} = \bm{x}_1' - \bm{x}_2', \qquad
\bm{p} = \frac12 \left( \bm{p}_1' - \bm{p}_2' \right),
\end{equation}
with $\bm{x}_i'$ and $\bm{p}_i'$ being the positions and momenta of the coalescing proton and neutron in the center of mass frame of produced deuteron. The oscillator frequency is determined by the charge root-mean-square radius of the deuteron, $\sqrt{\langle r^2 \rangle}_d = 1.96$~fm~\cite{CKL03a}, which leads to $\omega = 8.06 \times 10^{-3}$~GeV. We note that the above deuteron Wigner function reproduces very well that obtained in Ref.~\cite{CKL03a} using a more realistic deuteron wave function.

The calculated results for the $p_T^{}$ spectrum and elliptic flow of midrapidity ($|y|\le 0.5$) deuterons from the coalescence model are shown by dash-dotted lines in Fig.~\ref{fig:deut-pt} and Fig.~\ref{fig:deut-v2}, respectively. It is seen that except for small $p_T^{}$ the deuteron $p_T^{}$ spectrum in the coalescence model agrees almost perfectly with that in the transport model. The total deuteron number in this case is about $0.051$ and is only slightly larger than that from the transport model. For the deuteron elliptic flow $v_2^{}$, although the two models give similar values in the small $p_T^{}$ region, the one from the coalescence model is larger at higher $p_T^{}$ and the deviation between the two becomes larger  as $p_T^{}$ increases. As a result, the coalescence model does not describe the experimental data as well as the transport model. However, neither model could give a negative $v_2^{}$ for deuterons with $p_T^{} < 1.0$~GeV$/c$, contrary to that seen in the experimental data from the STAR Collaboration.

\section{Deuteron emission time distribution}
\label{sec:emission}

%%% FIG 6

\begin{figure}[t]
\centering
\includegraphics[width=0.85\columnwidth,angle=0,clip]{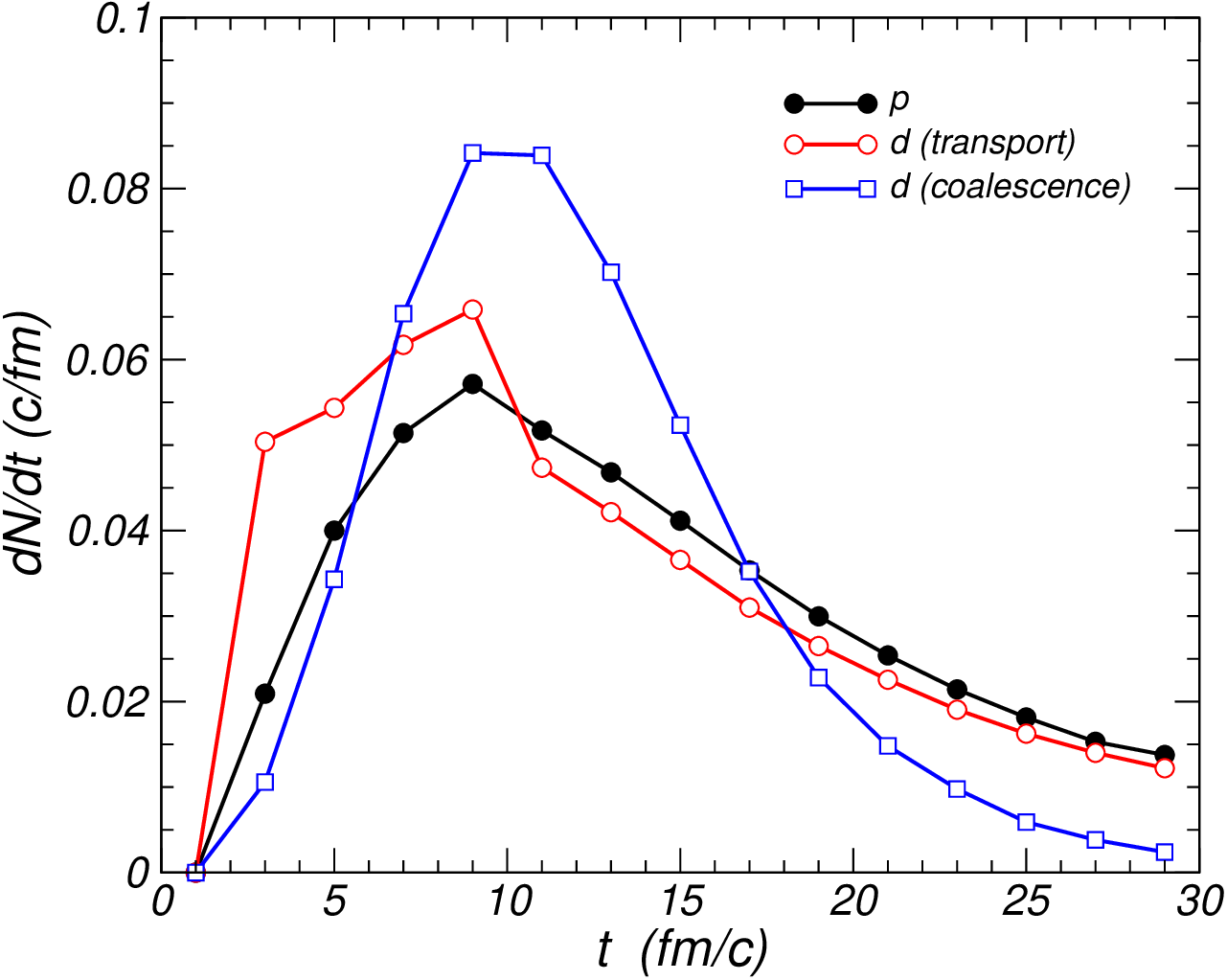}
\caption{(Color online) Normalized freeze-out time distributions of midrapidity protons (filled circles) and deuterons in the transport model (open circles) and in the coalescence model (open squares).} \label{fig:emit-time-coal}
\end{figure}

We have examined the distribution of the deuteron emission times, which  are the times for their last collisions in the transport model, and the results are shown in Fig.~\ref{fig:emit-time-coal}. The freeze-out times of nucleons are defined in the same way and their distribution is also given in Fig.~\ref{fig:emit-time-coal} for comparison. In the coalescence model, a deuteron is formed from a pair of freeze-out or emitted proton and neutron, so its emission time is given by the latest time of the two emitted nucleons. Results given in Fig.~\ref{fig:emit-time-coal} show that the distribution of deuteron emission times in the transport model is rather similar to that of nucleons. In the coalescence model, although the deuteron emission time distribution peaks at $t \sim 10$~fm/c, similar to that of nucleons, the two are otherwise quite different. In particular, there are more deuterons emitted at later times in the transport model than in the coalescence model. We note that the non-smooth deuteron early emission time distribution in the transport model is due to the incomplete destruction of initial deuterons, particularly those produced near the surface of the fire cylinder. Introducing a diffused hadron distribution in the initial state may help  smooth the deuteron emission time distribution during earlier times.

\section{Nucleon number scaling of deuteron elliptic flow}
\label{sec:scaling}

%% FIG 7

\begin{figure}[h]
\centering
\includegraphics[width=0.95\columnwidth,angle=0,clip]{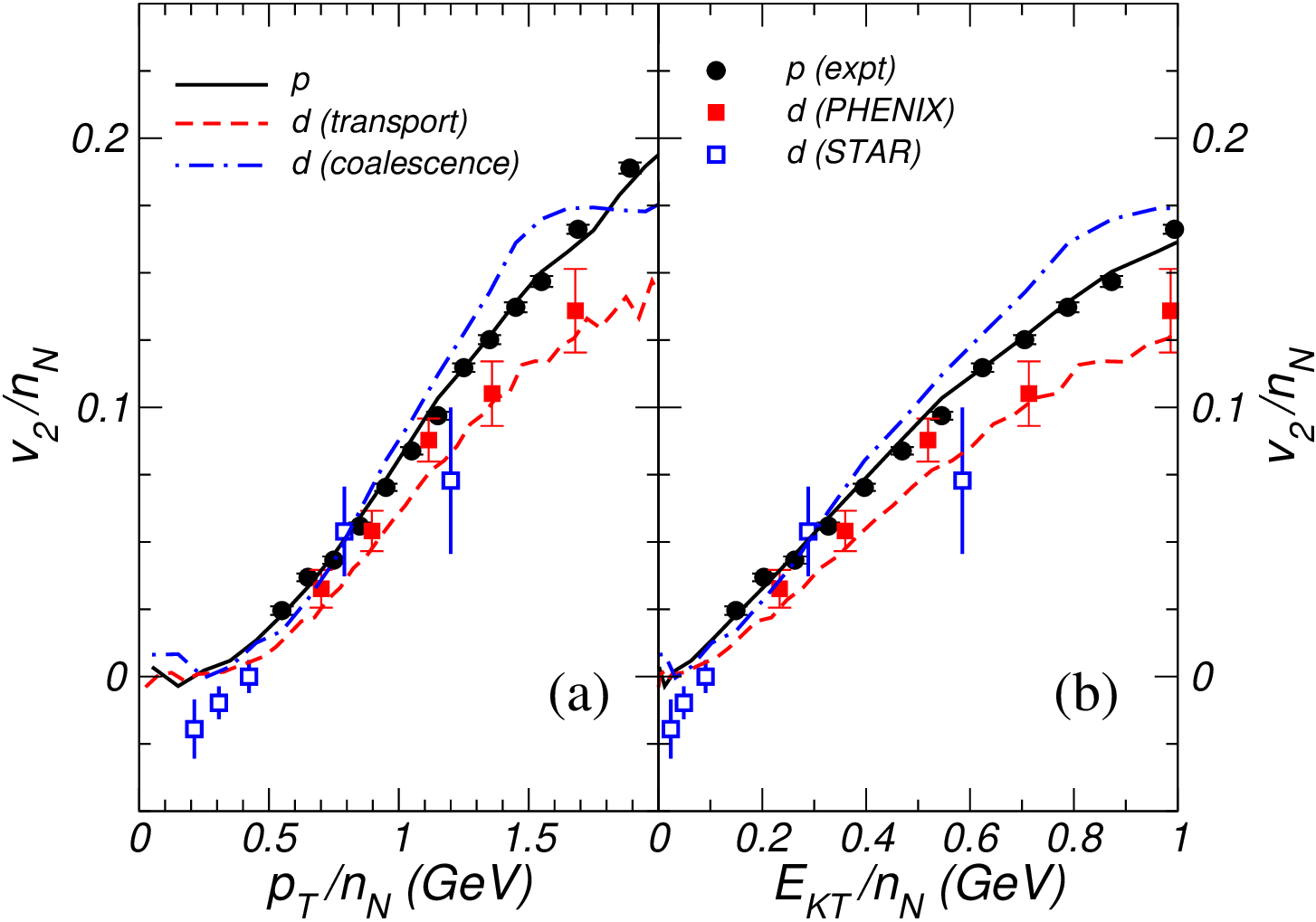}
\caption{(Color online) Scaled elliptic flow in (a) transverse momentum $p_T^{}$ and (b) transverse kinetic energy. The nucleon number $n_N^{}$ is $1$ for the proton and $2$ for the deuteron.} \label{fig:scaling}
\end{figure}

As in the case of the elliptic flows of identified hadrons, where a scaling according to the number of constituent quarks in a hadron has been observed~\cite{Greco:2004ex}, it is of interest to see if there is a similar scaling of the proton and deuteron elliptic flows according to their nucleon numbers. This is shown in Fig.~\ref{fig:scaling} for the elliptic flow per nucleon as a function of momentum per nucleon [panel (a)] or as a function of transverse kinetic energy [$E_{\rm KT}=(m^2+p_T^2)^{1/2}-m$] per nucleon [panel (b)]. It is seen that the nucleon number scaling of the proton and deuteron elliptic flows is not quite realized in the transport model, particularly at large $p_T^{}$.  This scaling is, on the other hand, very well satisfied in the results from the coalescence model, although they do not describe the data as well as the transport model.

\section{Summary and Discussions}
\label{sec:summary}

We have investigated deuteron production in heavy ion collisions based on a hadronic transport model. This is carried out by modifying the \textsc{art} code in the \textsc{ampt} model to include the production and annihilation of deuterons via the reactions $BB\leftrightarrow dM$ involving baryons and mesons as well as their elastic interactions with these hadrons. For the initial hadron distributions after hadronization of the quark-gluon plasma, we use the thermal distribution based on a blast wave model.

The initial hadrons are uniformly distributed in an elliptic shape with a spatial elliptic anisotropy of $-0.05$. With the parameters such as the initial temperature as well as the transverse flow velocity and anisotropy fitted to reproduce the measured $p_T^{}$ spectra and elliptic flows of protons, we have examined the production of deuterons in relativistic heavy ion collisions. The calculated deuteron $p_T^{}$ spectrum is comparable to, although somewhat softer than, the experimental data measured by the PHENIX Collaboration. The deuteron elliptic flow from the transport model also agrees reasonably with that measured by the PHENIX Collaboration. However, the deuteron elliptic flow was found to be positive at small $p_T^{}$ in contrary to the negative elliptic flow of deuterons observed by the STAR Collaboration.

We have also compared our transport model results with those of the coalescence model. While the two give very similar deuteron $p_T^{}$ spectra, the elliptic flows obtained in the two model are quite different, particularly at high $p_T^{}$. In addition, the coalescence model is found to give almost exact nucleon number scaling of the elliptic flow, while the transport model causes a deviation of the elliptic flow from the exact nucleon number scaling, which is, however, closer to the measured data. These results, including the non-negative values of the elliptic flow at very small $p_T^{}$, are not much changed if the cross sections for deuteron production and scattering are changed by a factor of two.

Our study has raised two interesting questions on deuteron production in relativistic heavy ion collisions. One is the relation between the transport model and the coalescence model for deuteron production, and the other concerns the negative deuteron elliptic flow at low $p_T^{}$. Since the binding energy of deuteron is small ($\sim 2.2$~MeV), the coalescence model has been considered an appropriate approach to describe deuteron production in nuclear reactions~\cite{Csernai:1986qf}. This is indeed the case in our study for the deuteron transverse momentum spectrum as the two models give very similar results. However, the deuteron elliptic flows from these two models are quite different, with the transport model giving a better description of the experimental data in comparison with the coalescence model. Part of this difference may be due to the neglect of rescattering of deuterons produced in the coalescence model. Although deuterons are produced from freeze-out nucleons in the coalescence model, they may undergo additional scattering in the hadronic matte, leading to a later emission time as results from the transport model have shown.  As to the negative deuteron elliptic flow observed in the preliminary experimental data, it is not seen in either the transport model or the coalescence model. The lack of negative deuteron elliptic flow in the coalescence model is expected as a negative deuteron elliptic flow requires a negative nucleon elliptic flow, which is not observed in the experimental data. In the transport model, a negative deuteron elliptic flow is possible if the final radial flow velocity is smaller than that reached in the present study, following the discussions in the end of Section~\ref{sec:deuteron1}. Whether this can still lead to a good description of the measured deuteron transverse momentum spectrum needs to be checked.   Further investigations both in theory and experiment are thus required to understand the elliptic flow of deuterons at low $p_T^{}$ and to shed more light on the mechanisms for the production of low $p_T^{}$ hadrons in relativistic heavy ion collisions.

\acknowledgments

This work was supported by the U.S. National Science Foundation under Grant No.\ PHY-0758115 and the Welch Foundation under Grant No.\ A-1358.

%%%%% Appendix %%%%

\appendix*

\section{Reactions for deuteron production, annihilation, and elastic scattering}\label{appendix:cross-section}

%%% FIG 8

\begin{figure}[t]
\centering
\includegraphics[width=0.95\columnwidth,angle=0,clip]{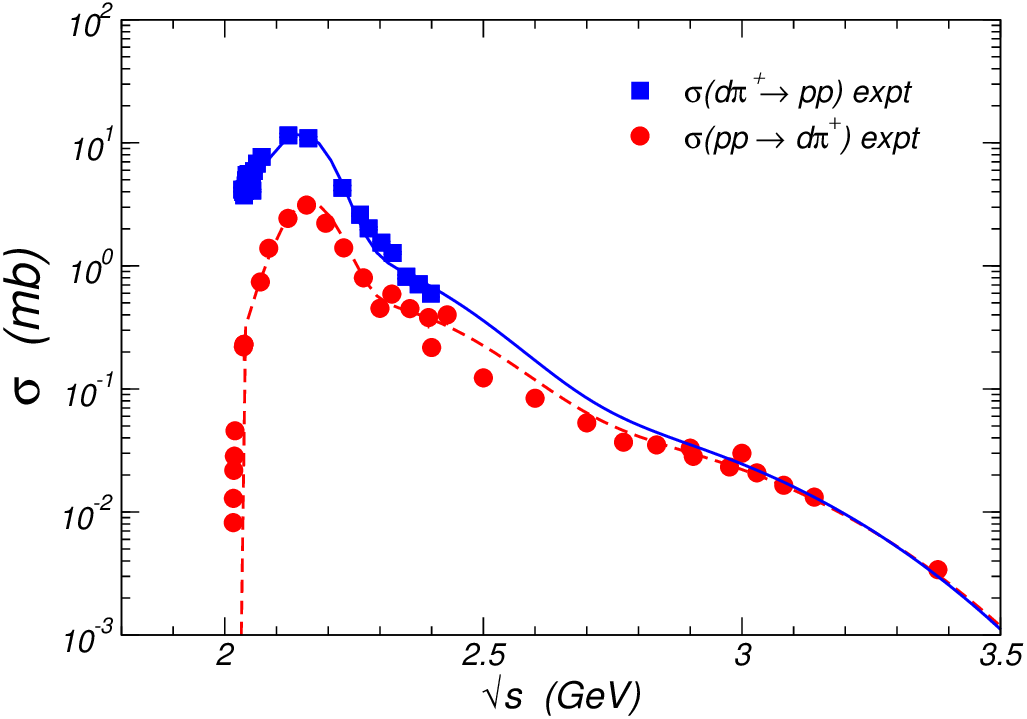}
\caption{(Color online) Experimental data on the total cross sections for $p p \to d \pi^+$~\cite{SKKS82,ALMD74} and for $d \pi^+ \to p p$~\cite{BGGG85,GGGG93,PAGO97}. The solid and dashed lines are the parametrization given in the text.} \label{fig:NN_dpi}
\end{figure}

The \textsc{art} code~\cite{LK95,LK96,LSZK01} implemented in the \textsc{ampt} code includes the interactions of $\pi$, $K$, $\eta$, $\rho$, $\omega$, $\phi$, $K^*$, $N$,
$\Delta(1232)$, $S_{11}(1535)$, $P_{11}(1440)$ as well as their antiparticles. To extend the model to include deuteron production and annihilation, we modify the \textsc{art} code to incorporate following reactions,
\begin{equation}
B B' \to M d, \qquad
M d \to B B',
\label{eq:scattering1}
\end{equation}
where $M = \pi$, $\rho$, $\omega$, $\eta$, and $B$ and $B'$ stand for baryons $N$,
$\Delta$, $P_{11}(1440)$, and $S_{11}(1535)$. For the cross sections of the reactions $BB'\to Md$, we assume that their angular integrated mean squared matrix elements that are averaged over initial and summed over final spins and isospins are the same as that for the reaction $N N \to d \pi$ at same center of mass energies. The cross sections for the inverse reactions $Md\to BB'$ are then determined from the detailed balance. Experimentally, the cross sections for both the reaction $p p \to d \pi^+$~\cite{SKKS82,ALMD74} and the reaction $\pi^+ d \to p p$~\cite{BGGG85,GGGG93,PAGO97} have been measured, and the former can be parameterized as
\begin{eqnarray}
\sigma(pp\to d\pi^+) = \frac14 \frac{p_\pi^{}}{p_N^{}} f(s),
\label{eq:pp}
\end{eqnarray}
where $p_N^{}$ and $p_\pi^{}$ are, respectively, the magnitude of the three-momenta
of initial and final particles in the center of mass frame. The function $f(s)$, which is proportional to the angular integrated mean squared matrix elements that are summed over initial and final spins for the reaction $p p \to \pi^+ d$, is given by
\begin{eqnarray}
f(s) &=& 26 \exp[-(s-4.65)^2/0.1] + 4 \exp[-(s-4.65)^2/2]
\nonumber \\ && \mbox{}
+ 0.28\exp[-(s-6)^2/10],
\end{eqnarray}
where the squared center of mass energy $s$ is in units of GeV$^2$ and $f(s)$ is in units of mb. For the inverse reaction $d \pi^+ \to pp$, its cross section is related to that for $pp\to d\pi^+$ via the detailed balance, i.e., $\sigma(d \pi^+ \to pp) = (2 p_N^2/3 p_\pi^2)\, \sigma(pp\to d\pi^+)$. These parameterizations are compared with the experimental data in Fig.~\ref{fig:NN_dpi}. The cross sections for the isospin averaged reactions $NN\to d\pi$ and $\pi d\to NN$ can then be obtained from $\sigma(NN\to d\pi)=\frac34\, \sigma(pp\to d\pi^+)$ and $\sigma(d\pi\to NN)=\sigma(d\pi^+\to pp)$.

%%% FIG 9

\begin{figure}[t]
\centering
\includegraphics[width=0.95\columnwidth,angle=0,clip]{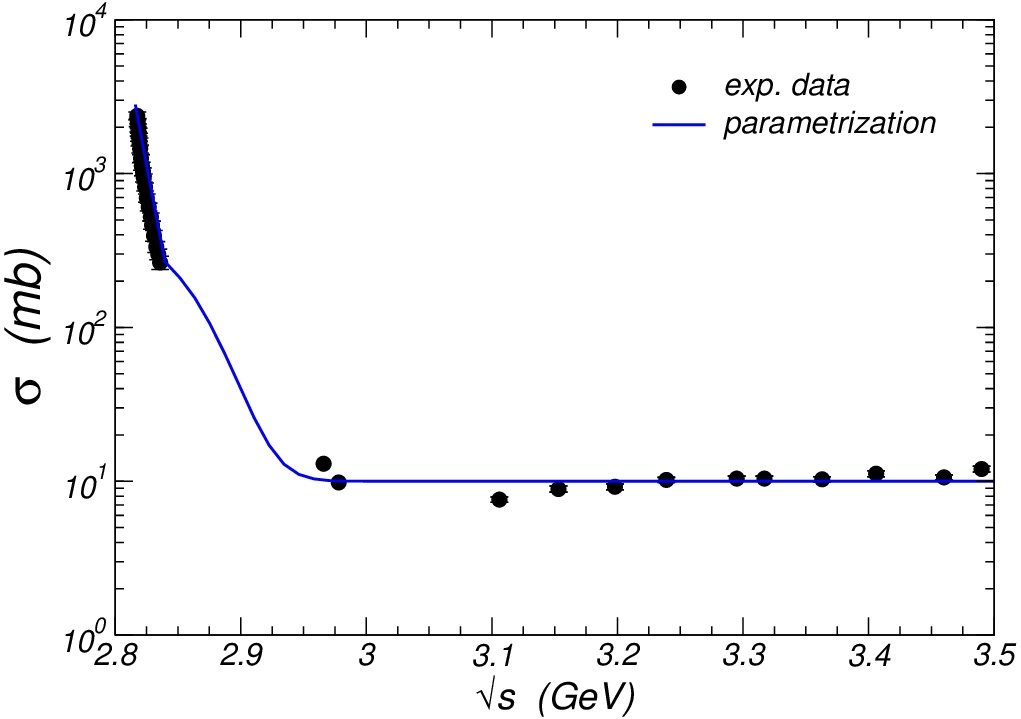}
\caption{(Color online) Experimental data from Refs.~\cite{SKDH83,KSTY85,HSHK02} on the total elastic cross sections for $N d \to N d$. The solid line is the parametrization used in this work.} \label{fig:ND}
\end{figure}

%%% FIG 10

\begin{figure}[t]
\centering
\includegraphics[width=0.95\columnwidth,angle=0,clip]{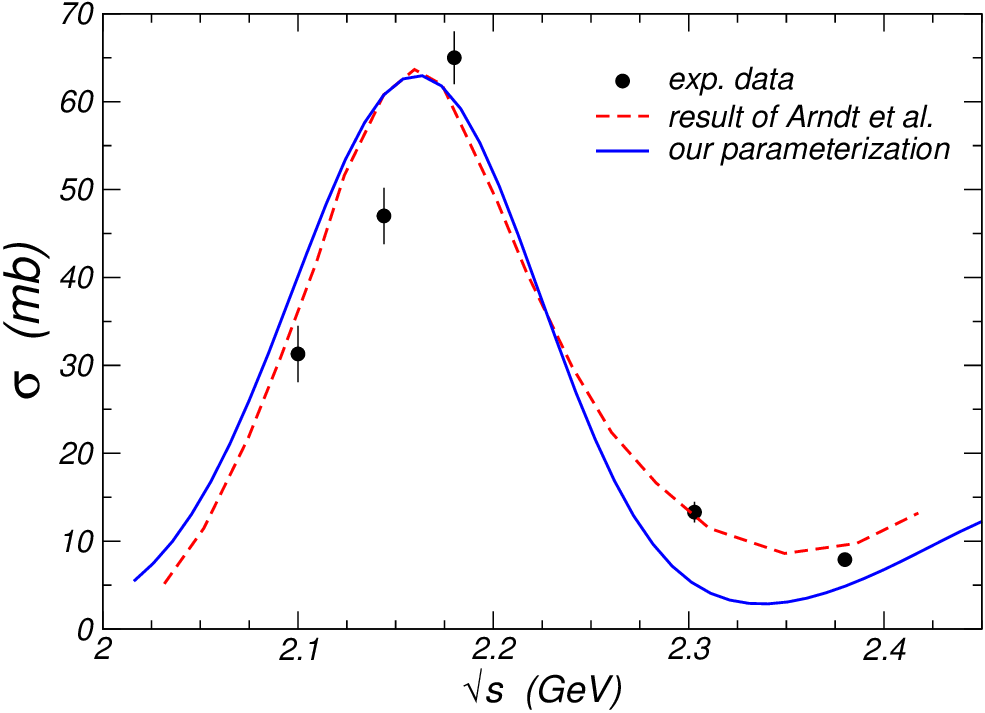}
\caption{(Color online) Experimental data from Refs.~\cite{BFK63,PFYFD63,Norem71,KDHM91,DKMM80} on the total elastic cross sections for $\pi d \to \pi d$. The solid line is the parameterization used in this work.} \label{fig:piD}
\end{figure}

In addition to the production and annihilation processes for deuterons, we also include their elastic scattering with mesons $M$ and baryons $B$,
\begin{equation}
M + d \to M + d, \qquad B + d \to B + d .
\end{equation}
As for the deuteron production cross sections from baryon-baryon reactions, the cross sections for deuteron elastic scattering with baryons and mesons are obtained from those for deuteron elastic scattering with nucleons and pions, respectively, assuming that at same center of mass energies they have same angular integrated mean squared matrix elements that are averaged over initial and summed over final spins and isospins.

For the empirical nucleon-deuteron elastic scattering cross section~\cite{SKDH83,KSTY85,HSHK02}, it can be parameterized by
\begin{eqnarray}
\sigma(Nd \to Nd) &=&
2500 \exp\left[-(s-7.93)^2/0.003 \right]
\nonumber \\ && \mbox{}
+ 300 \exp\left[-(s-7.93)^2/0.1 \right] + 10,
\nonumber \\
\end{eqnarray}
where the cross section is in mb and $s$ in GeV$^2$. The comparison with available experimental data are given in Fig.~\ref{fig:ND}.

The experimental data on the total cross sections for the $\pi d$ elastic scattering are very scarce. Therefore, we have to rely on phenomenological calculations. In Refs.~\cite{ASW94,OASW97}, based on a combined analysis of the reactions $pp \to pp$, $\pi d \to \pi d$, and $\pi d \to pp$, the helicity amplitudes of these reactions are determined which then leads to the total elastic scattering cross sections for $\pi d$ scattering. We use the results of
Refs.~\cite{ASW94,OASW97}, which can be parameterized as
\begin{eqnarray}
\sigma(\pi d \to \pi d) &=& 63 \exp\left[ -(s-4.67)^2/0.15 \right]
\nonumber \\ && \mbox{}
+ 15 \exp\left[ -(s-6.25)^2/0.3 \right],
\end{eqnarray}
where $s$ is in GeV$^2$ and the cross section is in mb. The comparisons with the results of Refs.~\cite{ASW94,OASW97} and with the data are given in Fig.~\ref{fig:piD}.

%\bibliographystyle{h-physrev4}
%\bibliography{biba-j,bibk-z}

\end{document}